# Deriving the Born Rule from a model of the quantum measurement process

Alan Schaum

*Abstract*

The quantum mechanics postulate called the Born Rule attributes a probabilistic meaning to a wave function. This paper derives the Born Rule from other quantum principles along with a model of the measurement process.

The nondeterministic nature of quantum measurements is hypothesized to arise from an ignorance of the quantum states of a measuring device's microscopic components. Their interactions with a system to be measured are modeled heuristically with any member of a particular class of stochastic processes, each of which generate the Born Rule. One member of the class appears particularly compatible with properties expected of quantum interactions.

## Background

> *Measurement in the Copenhagen interpretation remains an unexplained process, since there is nothing in the mathematics of quantum mechanics that specifies how or why the wave function collapses.* - *(*[1] p. 318)

In the early twentieth century new principles were being integrated with classical physics ideas in order to explain quantum phenomena. Orbital angular momentum was quantized in the Bohr model of the atom, then magnetic quantum numbers and electron spin were introduced, ultimately to be replaced by the self-consistent quantum mechanics of Heisenberg, Schrödinger, and Dirac. Electronic interactions with photons and the physical basis of the heuristic Pauli exclusion principle had to await the more general framework of quantum field theory for deeper explanations.

But one axiom appended to early quantum principles, the Born Rule, continues to lack a consensus explanation after 100 years. This paper offers a less exotic explanation than most, describing how the effect of a quantum measurement can be modeled as a series of simple interactions of a quantum system with the components of the measuring device.

## Introduction

First a general model is introduced to describe the "collapse" of a quantum mechanical wave function that a measurement induces. Then a more detailed class of associated mathematical processes is introduced that explains how a mixed quantum state S could be transformed by a measuring device D into an eigenstate of the observable O that the device is designed to measure.

In classical physics no general definition of measurement seems necessary. One knows it when one sees it. But more is required in quantum mechanics, given the mystery of what happens to a wave function during a measurement. Quantum mechanics predicts that some systems can exist as superpositions of discrete eigenstates of O, and because a conventional measurement can produce only one of the associated eigenvalues, it is generally agreed that the act of measuring transforms a mixed quantum state into an eigenstate. It also seems incontrovertible that:

I. A measurement of an observable O is always made by a *macroscopic* device D.
II. Associated with D is a preferred basis in Hilbert space, the eigenvectors of O.
III. D reports any pure state faithfully and never reports a state absent from S.

The evolution of S into an eigenstate is governed by probabilities and so should be interpretable as a stochastic process. In the model proposed here, D's microscopic components perturb S serially according to any one of a well-defined class of stochastic processes. The perturbations produce a series of complex rotations (unitary transformations) of the Hilbert-space vector associated with S. The models all predict that every such process ends in some eigenstate of O, and with probabilities defined by the Born Rule. Those eigenstates also define the basis in which the stochastic models are most easily formulated.

The effect of a measurement is described better as a random realignment of the quantum state vector, which remains normalized, rather than a collapse (as in a projection). Thus the models preserve vector length at every step. Also, like the Born Rule itself, the models do not relate directly to any particular physical mechanism of measurement. The hope is that, for any particular type of device, ultimately some member of the proposed class of models will be recognized as consistent with a more focused quantum mechanical description of the associated measurement process. However, one member of the class of model processes (defined by equation (12)) seems particularly appropriate as a general candidate.

The quantum state S initially consists of a superposition of some $M$ eigenstates of O with amplitudes whose squared magnitudes (here called intensities) are perturbed sequentially upon each encounter with a component of D. Each encounter is in turn described by a sequence of more elementary binary processes, in which a small intensity amount $\epsilon$ is removed from a randomly chosen eigenstate, and then is recaptured by a random eigenstate. The many microscopic components of the macroscopic D accommodate a potentially large number of such encounters ($\sim 10^{23}$ if necessary). The idea of modeling these interactions with S as a sequence of small changes is motivated by considering a measuring device that has been reduced in complexity down to a single molecule, which can be expected only to *perturb* a quantum state S, not necessarily transform it into an eigenstate.

## **Elementary stochastic process**

The initial $M$ intensity values are modeled as integers $a_i$ times $\epsilon$, with $\epsilon = 1/N$, where $N$ is a large integer and

$$\sum_{i=1}^{M} a_i = N \,. \tag{1}$$

The state S is represented here as S $= (a_1, a_2, \ldots, a_M)\epsilon$, not the usual Hilbert space vector, because the complex phases of the wave function do not figure in the models. The condition of equation (1) is maintained throughout the progression of S to its final pure eigenstate.

The evolution of intensity $a_i$ proceeds as follows. Some model transition probability distribution $p_M(a_i)$ is defined that obeys the constraints:

$$\sum_{i=1}^{M} p_M(a_i) = 1,\ p_M(0) = 0,\ p_M(a_i) \neq 0 \ if \ a_i \neq 0 \,. \tag{2}$$

A transaction between two randomly selected intensities is initiated in which an amount $\epsilon$ can be exchanged, according to the following process. With probability $p_M(a_i)$, intensity $\epsilon a_i$ donates an amount $\epsilon$ to a repository. Or with probability $\sum_{j=1, j\neq i}^{M} p_M(a_j) = 1 - p_M(a_i)$, some other intensity donates $\epsilon$. Regardless of the source of the donation, $\epsilon a_i$ re-acquires $\epsilon$ with probability $p_M(a_i)$, and some other intensity acquires it with total probability $1 - p_M(a_i)$.

The second equality in conditions (2) is required by the general model, because once an intensity reaches zero, it can no longer contribute to the repository. That equality then also insures that once a component exits the evolution process, it will not return. This effects a "stopping rule" inherent to the process. Once $a_i$ reaches zero, it stays there. Also, during the evolution of S, with probability one some $a_i$ will reach zero (at least in the limit of an infinite number of interactions). Thereafter the process continues until only one nonzero intensity survives. The inequality in condition (2) insures that a nonzero component of the evolving S cannot defect from the evolution process. Thus (2) guarantees requirement III above.

Note that the evolution of intensity $\epsilon a_i$ is governed by only $p_M(a_i)$ and is not dependent on the individual probabilities $p_M(a_j), j \neq i$. Therefore, in the stochastic process of $a_i$ evolving into either $N$ or 0, the labels on the intensities where the other $N - a_i$ $\epsilon$-intensities reside have no effect. Similarly, binary exchanges occurring amongst those eigenstates do not effect the probabilistic fate of intensity $a_i$. Consequently, the probability $P_i(a_1, \ldots, a_i, \ldots, a_M)$ that $a_i$ will evolve asymptotically to the value 1 has the same value as it would in a competition with a pooled version of all other intensities. Thus, the $M = 2$ result can be used to predict the more general result, and the probability that the $i^{th}$ eigenstate is reached can be expressed as

$$P_i(a_1, \ldots, a_i, \ldots, a_M) = P_1(a_i, N - a_i)\,. \tag{3}$$

$\underline{M = 2}$

To find the probabilities for the case $M = 2$, define $P(a) \equiv P_1(a, N - a)$, the probability that the initial state $S = (a, N - a)\epsilon$ evolves into the pure eigenstate $(1, 0)$. The stochastic evolution of the first intensity $\epsilon a$ is governed at each step according to various transition possibilities:

$$\begin{aligned} P(a) = p_M(a)\big((p_M(a))P(a) + (1 - p_M(a))P(a-1)\big) \\ +(1 - p_M(a))\big(p_M(a)P(a+1) + (1 - p_M(a))P(a)\big) \end{aligned} \tag{4}$$

The first line corresponds to where $\epsilon$ can land if $\epsilon a$ is the donor, the second if the other intensity donates to the repository.

Dividing equation (4) by $p_M(a)(1 - p_M(a))$ yields the surprising result:

$$P(a+1) = 2P(a) - P(a-1), \tag{5}$$

which is independent of $p_M$ ! The division is always allowed because, according to (2), either divisor is zero only when the evolution process for $a$ terminates, at either $a = 0$ or $a = N$.

Because equation (5) requires that second-order differences of $P(a)$ be zero, its general solution has the form $P(a) = \alpha + \beta a$. The boundary conditions $P(0) = 0$ and $P(N) = 1$ then yield $P(a) = a/N = a\epsilon = P_1(a, N - a)$. From equation (3) it then follows that

$$P_i(a_1, \ldots, a_i, \ldots, a_M) = a_i \epsilon \; . \tag{6}$$

This is the Born Rule: The probability that the state $S$ evolves into its $i^{th}$ eigenstate equals that eigenstate's initial intensity. Thus, the Rule is satisfied by the above measurement model in combination with a family of associated stochastic processes, constrained only by equation (2).

## **Suggested experimentation**

"The great tragedy of science–the slaying of a beautiful hypothesis by an ugly fact."
-Thomas Henry Huxley

The main premise of this paper, that the measurement process can be modeled as a series of perturbative interactions, is falsifiable, in principle. As a test of the premise, components of a measurement device could be pared down to minimalist essentials to reduce the number of hypothesized interactions, producing a new kind of partial measurement. The surviving modified version of S produced by such a device could be examined for evidence of the interruption of an evolutionary process such as modeled above. It could produce a final state that is neither the original nor an eigenstate.

One can envision, for example, Stern-Gerlach-type experiments [2] in which horizontally polarized spin-1/2 silver atoms (corresponding to the $M = 2$ case) impinge on a modified device (see figure). A second half-sized detection screen with a hole larger than the upper

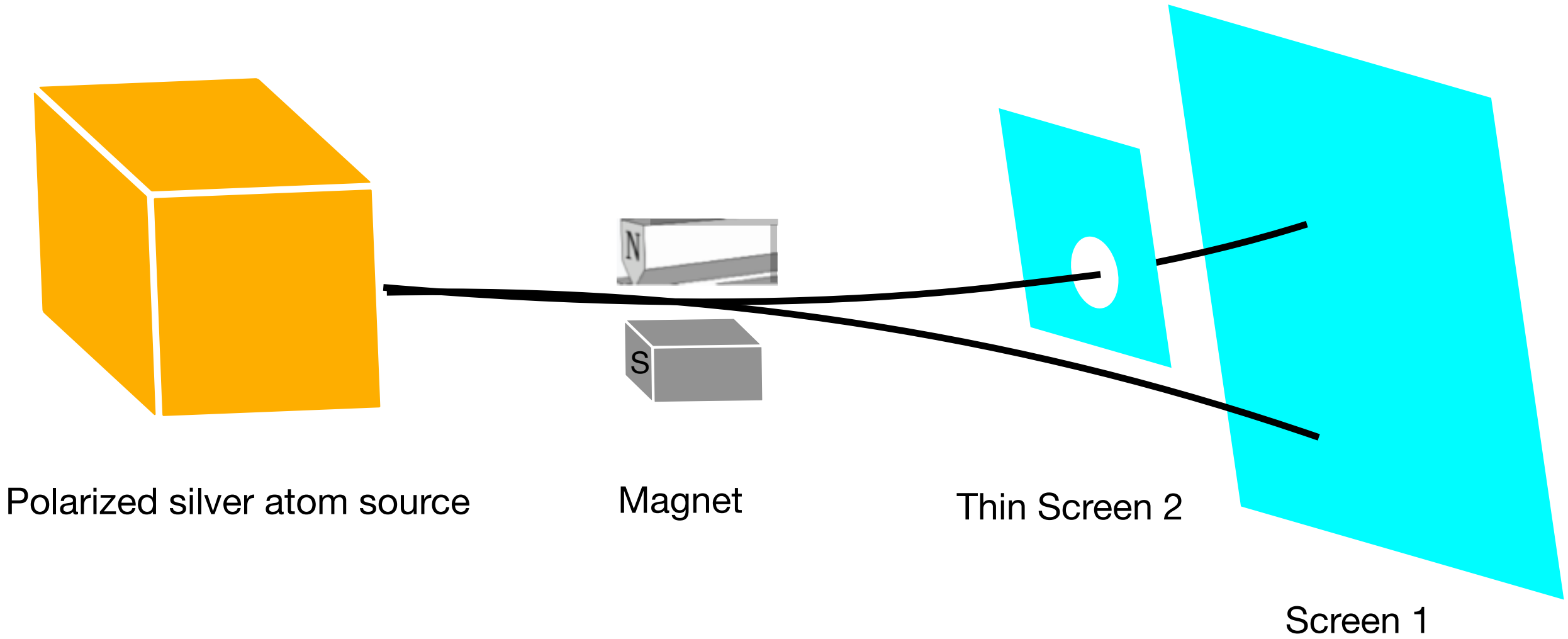

**Notional Modified Stern Gerlach Apparatus**

beam width is first inserted between the magnet and the original screen, an arrangement that should not disrupt the original experiment, with half spin up and half spin down detections on screen 1. If the hole is then filled, the beam detected by screen 1 is polarized with spin down. But if instead a thin enough film of the detection material can be inserted into the hole, the above model predicts that the original quantum state S of the silver atoms could be converted into other mixed states, and screen 1 could be used to collect statistics to test that hypothesis.

The feasibility of the test relies on certain contingencies. For example, there must be a large enough number of interactions with small $\epsilon$'s to enable conversion of S into a measurable mixed state. It should be noted also that in general the number of elementary binary exchanges corresponding to the interaction of S with a single component of D might vary with device, as well as during the entire measurement process. The mean values computed below refer to the total number of binary interactions over all detector components required for conversion from a superposition of eigenstates into a pure eigenstate.

**Number of steps to complete the process**

As in the above argument generalizing the $M = 2$ probabilities to $M \geq 2$, the mean number of interactions $v_{a_i}$ for a state whose $i^{\text{th}}$ component is $a_i$ to evolve into completion, that is, into a value of either 0 or $N$, can be computed by considering the $M = 2$ case. Then, as in equation (4), four transitions can occur, starting from the value $a_i$ $(\equiv a)$, but now each adds one interaction to the process. Therefore,

$$v_a = p(a)\big((p(a))v_a + (1 - p(a))v_{a-1}\big) + (1 - p(a))\big(p(a)v_{a+1} + (1 - p(a))v_a\big) + 1\ , \qquad (7)$$

where it has been assumed that the probabilities in equation (2) are independent of $M$, which ultimately must change in the evolution process, ending in the value one. Equation (7) can be rewritten:

$$v_{a+1} = 2v_a - v_{a-1} - q_a\ , \quad \text{with} \quad q_a = \Big[p(a)\big(1 - p(a)\big)\Big]^{-1} \ \ (0 < a < N)\ . \qquad (8)$$

Next let $d_a = v_a - v_{a-1}$. Equation (8) becomes $d_{a+1} = d_a - q_a$, whose solution is $d_a = d_1 - \sum_{i=1}^{a-1} q_i$, because $v_0 = 0$. Because also $d_1 = v_1$, this means that

$$v_a - v_1 = \sum_{n=2}^{a} d_n = \sum_{n=2}^{a}\left[v_1 - \sum_{i=1}^{n-1} q_i\right] = (a-1)v_1 - \sum_{n=2}^{a}\sum_{i=1}^{n-1} q_i\ .$$

On reversing the summation order, this simplifies to: $v_a = a v_1 - \sum_{i=1}^{a-1} q_i(a - i)$, $[a \geq 2]$.

To evaluate $v_1$, use the boundary condition $v_N = 0$. The final closed form for the mean time to completion from a state with initial intensity $a\epsilon$ is then

$$v_a = \frac{a}{N}\sum_{i=1}^{N-1}(N-i)q_i - \sum_{i=1}^{a-1} q_i(a-i),\ \ [a \geq 2]\ ; \qquad v_1 = \frac{1}{N}\sum_{i=1}^{N-1}(N-i)q_i\ ; \quad v_0 = 0\ . \tag{9}$$

Notice that $q_0$ does not appear, and therefore neither does $p(0)$, so that according to equation (2), $q_i$ is well-defined in equation (9).

Each step in the process involves two intensities, so that the total mean time for completion of the stochastic evolution of S $= (a_1, \ldots\, a_i, \ldots, a_M)\epsilon$ is

$$\frac{1}{2}\sum_{i=1}^{M} v_{a_i}\ . \tag{10}$$

It should be noted that the number of steps counted here includes those in which no change occurs in any of the intensities. According to equation (4), these "null" interactions occur a fraction $p^2(a_i)$ of the time for intensity $a_i$, which can vary with each iteration if $p$ depends on $a_i$.

The above analysis also describes probability distributions $p_M(a_i)$ for which $M = 2$, because $M$ changes only at the last step of the process. For the particular "uniform" probability model $p_2(a_i) = \left(\frac{1}{2}\right)(1-\delta^0_{a_i})(1-\delta^N_{a_i}) + \delta^N_{a_i}$ (with $\delta$ the Kronecker delta), it follows that $q_1 = q_2 = 4$, and the sums in equation (9) are computable, with the result $v_{a_i} = 2a_i(N-a_i)$.

Again, because both eigenstates participate in each interaction, the mean number of binary interactions until a pure eigenstate arises is $\frac{1}{2}\sum_{i=1}^{2} 2a_i(N-a_i)$. With equation (1) this simplifies to $2a_1a_2$. Notice that null interactions occur in the $M = 2$ model during an average of $p^2(a_1) + p^2(a_2) = \frac{1}{2}$ of the transitions, and so the mean number of *nontrivial* transitions is only $a_1a_2$.

Note finally that once a final eigenstate has been reached, the stochastic process can be thought of as continuing indefinitely with repeated null interactions, as the final state continues to be "measured" and remains pure.

**Number of nontrivial steps in the process**

If the above model is ever tested experimentally, it could be useful to know the mean number of steps involved in the evolution of S that involve actual changes in intensities. According to equation (7), given that the value of $a$ changes, it has equal probabilities of increasing or decreasing by one. Therefore, the mean number $w_a$ of non-null steps involved in the evolution of intensity $a$ satisfies

$$w_a = \frac{1}{2}w_{a-1} + \frac{1}{2}w_{a+1} + 1,\ (0 < a < N),$$

which can be written

$$w_{a+1} = 2w_a - w_{a-1} - 2.$$

This was considered earlier (equation (8)) and is solved in equation (9), resulting (for $q_a = 2$ ) in $w_a = a(N - a)$. Then, as in equation (10), the mean number of non-trivial steps in the total process of evolution to the final state is

$$\frac{1}{2}\sum_{i=1}^{M} w_{a_i} = \frac{1}{2}\sum_{i=1}^{M} a_i(N - a_i) = \sum_{i<j}^{M} a_i a_j \ , \tag{11}$$

which is maximized at the value $\dfrac{(M-1)N^2}{2M}$ when $a_i = N/M$. So, $O(N^2)$ steps with changes in intensity might be necessary for evolution to an eigenstate.

## **<u>Not so spooky?</u>**

The proposed experiment and model resurrect the knotty question of “spooky action at a distance.” Interactions with the atom’s wave function at the top of one screen affect the amplitude of the wave function at the bottom of another. However, the mathematical representation of the spin-1/2 atom’s quantum state includes two separate wave functions, one for the spin up and one for the down component. That is, its mathematical representation is as an outer product of a continuous spatial representation with a two-component spin representation. The change in one component caused by the measuring screen is effected by a change in the other component across the spin direction, not the spatial. The transformation is achieved mathematically by a 2x2 unitary matrix operating in spin space. Thus, the distance across which the action occurs may seem spooky, but it is transverse to the spatial representation, and (as is commonly acknowledged) no problem arises with relativistic causation.

## **<u>A historical precedent for sleight of hand</u>**

*“God made the integers. All else is the work of man.” - Leopold Kronecker*

Max Planck regarded his seminal paper [3] in the year 1900 as an “act of desperation” [4]. He was motivated by his belief that “a theoretical interpretation (of his formula for the blackbody spectrum) had to be found at any cost” [ibid]. For many years after, he suspected that his method of quantizing the interactions between a blackbody cavity’s wall and its radiation field had “no more than a formal significance” [5]. “He did not believe (the energy of radiation) was really chopped up into quanta” ([1] p. 26). He and his contemporaries “all believed that it was nothing more than the usual theorist’s sleight of hand, a neat mathematical trick on the path to getting the right answer” ([1] p. 27). His fictional contrivance, the quantization constant $h$, could be adjusted to account for the experimental data, but he did not understand the truth behind his heuristic model of the physics.

By contrast, in this paper the limit $\epsilon \to 0$ can be taken and still give the desired result, the Born Rule. However, if the underlying premise of this model is necessary, then $N = 1/\epsilon$ is finite, because the number of components of a measuring device is finite. This would mean that $\epsilon$ cannot be zero.

Planck found the fundamental quantum constant, which led to quantized physical entities, such as angular momentum and energy. The various charges and masses of elementary particles are also quantized. Theories of quantum gravity even raise the possibility of space and time being quantized. In the above model, $\epsilon$ represents a small fraction of a probability, which is a pure number. A pure number can be derived from the fundamental physical constants, if one includes a mass estimate of the lightest known massive elementary particles. The neutrino/Planck mass ratio is thought to be nonzero but $< 10^{-29}$, perhaps an upper limit to the value of $\epsilon$. Might $\epsilon$ reflect the quantization of number itself, when applied to the physical world? Could $\epsilon$ underly all other quantizations?

## **Conclusion and Future Prospects**

"Plurality should not be posited without necessity." - William of Occam

A simple explication of the quantum measurement process as a series of elementary perturbative interactions appears supportable. Exotic semi-metaphysical explanations such as the "many worlds interpretation" [6] of quantum mechanics seem unnecessary. Nor is there a need to give up on notions of objective reality. The moon is there, even if nobody is looking at it, because it is macroscopic and constantly measures itself, or "decoheres."

It appears that the probabilistic nature of a quantum measurement can be understood to arise merely from an incomplete knowledge of the detailed microscopic interactions between the system under study and the components of the measuring device. Analogous to the situation with a classical gas. Quantum measurements give differing results despite "identical" conditions, because no experiment has ever actually been repeated. Only the macroscopic, not the quantum condition of the measurement device is replicated.

The models of this paper share with the Born Rule their generality. None explores the detailed quantum processes of any specific type of measurement device. A synopsis of the main assumption and reasoning:

1. Anything qualifying as a measurement device has many microscopic components that individually perturb the quantum system being measured.
2. Associated with every quantum measurement is a probabilistic mechanism that transforms any superposition of eigenstates into some pure eigenstate.
3. Therefore, the course of a measurement must be describable as a stochastic process.

This paper has shown that some of the simplest, pairwise exchange models of such processes, consistent with general quantum principles, are also governed by the Born Rule. Coincidence perhaps. But this property could inform future attempts at more detailed physics explanations for any particular measurement modalities.

Also, the demonstration that an entire class of stochastic models obeys the Born Rule raises the prospect that the Rule's universality might be associated not with a single transition probability model $p_M(a)$, but rather with a subset of those belonging to some class, such as defined by equation (2).

However, the choice

$$p_M(a_i) = \epsilon a_i \tag{12}$$

seems intuitively attractive from a general physics perspective. It says that the stronger the presence of an eigenvector in a quantum system, the stronger will be the degree of its interaction with a measurement device.

Finally, it can be argued that these ideas at most replace the question of why a measurement should obey the Born Rule (equation 6) with why transition probabilities should (perhaps) obey the analogous equation (12). But at least the Rule's demonstrated consistency with the above microscopic measurement model might serve as the touchstone to a deeper truth.

And it has not escaped notice that the specific underlying model postulated here immediately suggests possible implications for nascent quantum technologies, such as in computing and cybersecurity.